\documentclass[12pt]{article}  
\usepackage{graphicx}

\def\R{ {\rm R \kern -.31cm I \kern .15cm}}
\def\C{ {\rm C \kern -.15cm \vrule width.5pt \kern .12cm}}
\def\Z{ {\rm Z \kern -.27cm \angle \kern .02cm}}
\def\N{ {\rm N \kern -.26cm \vrule width.4pt \kern .10cm}}
\def\1{{\rm 1\mskip-4.5mu l} }
\def\lsim{\raise0.3ex\hbox{$<$\kern-0.75em\raise-1.1ex\hbox{$\sim$}}}
\def\gsim{\raise0.3ex\hbox{$>$\kern-0.75em\raise-1.1ex\hbox{$\sim$}}}
\def\noi{\noindent}

\def\beq{\begin{equation}}   \def\eeq{\end{equation}}
\def\bea{\begin{eqnarray}}  \def\eea{\end{eqnarray}}
\def\nn{\nonumber}
\def\noi{\noindent}
\def\beeq{\begin{eqnarray}} \def\eeeq{\end{eqnarray}}

\begin{document}

\begin{center} {\large \bf } \vskip 2 truemm

 {\large \bf Some Remarks on Effective Range} \vskip 2 truemm
 {\large \bf  Formula in Potential Scattering}

 \par \vskip 5 truemm

{\bf Khosrow Chadan }\\ {\it Laboratoire de Physique
Th\'eorique}\footnote{Unit\'e Mixte de Recherche UMR 8627 - CNRS}\\   
{\it Universit\'e de Paris XI, B\^atiment 210, 91405 Orsay Cedex,
France}\\
{\it  (Khosrow.Chadan@th.u-psud.fr)} \par \vskip 5 truemm
\end{center}
\vskip 1 truecm

\begin{abstract}
In this paper, we present different proofs of very recent results on the necessary as well as sufficient conditions on the decrease of the potential at infinity for the validity of effective range formulas in 3-D in low energy potential scattering (Andr\'e Martin, private communication, to appear. See Theorem 1 below). Our proofs are based on compact formulas for the phase-shifts. The sufficiency conditions are well-known since long. But the necessity of the same conditions for potentials keeping a constant sign at large distances are new. All these conditions are established here for dimension 3 and for all angular momenta $\ell \geq 0$.
\end{abstract}

\vskip 3 truecm 

\begin{flushleft}
LPT Orsay  08-41\\
November 2008
\end{flushleft}
 
\newpage
\pagestyle{plain}
\baselineskip 20pt
\noi {\bf I - \underline{Introduction}} \\

We consider potential scattering in 3-dimensions with a real, local, and spherically symmetric potential $V(r)$ [1-3]. The spherical symmetry allows us to consider separately each angular momentum state $\ell$, and study the corresponding radial Schr\"odinger equation. We begin with the $S$-wave ($\ell = 0$) in order to see clearly the main points without algebraic complications. We therefore have the following approximation, called effective range formula, for the $S$-wave phase-shift $\delta_0 (k)$ [1-3]~:

\beq
\label{1e}
k \ {\rm cotg}\  \delta_0 (k) \ \mathrel{\mathop {\cong}_{\hbox{$k$ small }}} \ - {1 \over a_0} + {1 \over 2} \ r_0k^2 \ , 
\eeq

\noi where $k$, in appropriate units, is the momentum of the particle. This formula was found very useful in the past for low energy scatterings of particles, especially for nucleon-nucleon scattering. The conditions for the validity of (\ref{1e}) are not very stringent. In general, for having a decent scattering theory, and finite phase-shifts $\delta_{\ell} (k)$, continuous and bounded for all $\ell \geq 0$, and all $k \geq 0$, one needs only [1-3]

\beq
\label{2e}
r V(r) \in L^1(0, \infty )\ .
\eeq

\noi Under this condition, one has then, at most, a finite number $n$ of bound states, and the Levinson Theorem [1-3]~:

\beq
\label{3e}
\left \{ \begin{array}{l} \delta_0 (0) - \delta_0 (\infty ) = n \pi \ \hbox{if no resonance at $E = k^2 = 0$\ ,} \\ \\ \delta_0 (0) - \delta_0 (\infty ) = n \pi + {\pi \over 2} \ \hbox{a resonance at $E = 0$\ .}\end{array} \right . 
 \eeq

Usually, one chooses $\delta_0 (\infty ) = 0$. In the case of a resonance at $E = 0$, $a_0$ becomes infinite. We shall assume (\ref{2e}) throughout this paper.\par

Whatever the number of bound states $n$, one has then \cite{3r,4r}~:\\

a) for having a finite scattering length $a_0$, it is sufficient to have

\beq
\label{4e}
r^2 V(r) \in L^1(1, \infty )\ ;
\eeq
\vskip 5 truemm

b) and for having a finite effective range $r_0$ it is sufficient to have
\beq
\label{5e}
V(r)  \ \mathrel{\mathop {\sim}_{r \to  \infty}} \ r^{-s}\ , \quad s > 5 \ .
\eeq
\vskip 5 truemm

\noi {\bf Remark 1.} Assuming only (\ref{2e}), the phase-shifts $\delta_{\ell}(k)$ are defined only for real values of $k$, and without extra conditions on the decrease of the potential at infinity, they cannot be extended to complex values of $k$. However, if one assumes $e^{2\mu r} V(r) \in L^1 (1 , \infty )$, then the $S$-matrix $S_{\ell}(k) = \exp [2 i \delta_{\ell} (k)]$ is meromorphic in $|{\rm Im}\ k| < \mu$, and one can extend (\ref{1e}) in this strip of the $k$-plane, which may contain the lowest bound states $k_j = i\gamma_j$, $\gamma_j < \mu$. This is the case for neutron-proton scattering. We refer the reader to \cite{3r} for more details.\\

Very recently, the above conditions have been made more precise (Andr\'e Martin, private communication, to appear). His essential points can be summarized as follows~: \\

\noi \underline{\bf Theorem 1.} For the $S$-wave, and in the absence of a resonance at $E = k^2 = 0$, if the potential $V(r)$ keeps a constant sign beyond some finite $R$, then for having $a_0$ finite, it is both necessary and sufficient to have $r^2 V(r) \in L^1(0, \infty )$. Then, in order to have also $r_0$ finite, it is again necessary and sufficient to have $r^4V(r) \in L^1(0, \infty )$.  \\

One may ask then what happens to the second term in (\ref{1e}) when $r^4 V(r) \notin L^1 (1, \infty )$~? This also has been studied for $V(r) \sim r^{-s}$ at infinity \cite{4r}, and shown by Andr\'e Martin on examples.\\

\noi {\bf Remark 2.} If the effective range formula (\ref{1e}) is exact for all $k$, the corresponding potential in decreasing exponentially at infinity \cite{5r}. It corresponds to a Jost function analytic in the whole $k$-plane, with one zero and one pole.\\

The purpose of the present paper is to give a different proof of Theorem 1, and generalize also to all anguler momenta $\ell \geq 0$.\par

We end this introduction by quoting the following general theorem, which will be useful for our proofs~:\\

\noi \underline{\bf Theorem 2. (Hille, \cite{6r}).} Consider the differential equation

\beq
\label{6e}
\phi''_0 (r) - V(r) \phi_0 (r) = 0 \ .
\eeq

\noi If $rV(r) \in L^1(0, \infty )$, this equation has a unique solution $\chi_0 (r)$ such that

\beq
\label{7e}
\lim_{r \to \infty} \chi_0 (r) = 1 \quad , \quad \lim_{r \to \infty} \chi '_0 (r) = 0 \ . 
\eeq

\noi If $V(r)$ is real, and ultimately keeps a constant sign, the condition on $V$ is both necessary and sufficient for the existence of such a solution $\chi_0$. There is also a solution $\phi_0(r)$, non-unique (modulo the addition of $\alpha \chi_0 (r)$), such that

\beq
\label{8e}
\lim_{r \to \infty} {\phi_0 (r) \over r} = 1 \quad , \quad \lim_{r \to \infty} \phi '_0 (r) = 1 \ .
\eeq

\noi If $V(r)$ satisfies also the condition $r^2 V(r) \in L^1 (0, \infty )$, then $\chi_0 (r)$ satisfies

\beq
\label{9e}
\lim_{r \to \infty} r \left [ \chi_0 (r) - 1 \right ] = 0 \quad , \quad \lim_{r \to \infty} r^2 \chi '_0 (r) = 0\ ,
\eeq

\noi and there exists a unique solution $\psi_0 (r)$ such that

\beq
\label{10e}
\lim_{r \to \infty}  \left [ \psi_0 (r) - r \right ] = 0 \quad , \quad \lim_{r \to \infty} r \left [ \psi '_0 (r)  - 1 \right ] = 0\ .
\eeq

\noi Again, if $V(r)$ keeps a constant sign, the condition $r^2V(r) \in L^1 (1, \infty )$ is both necessary as well as sufficient for the existence of such a solution $\psi_0$. \par

Obviously, $\phi_0$ and $\chi_0$, or $\psi_0$ and $\chi_0$, are two independent solutions of (\ref{6e}) since the Wronskians at $r = \infty$, are, according to (\ref{7e}), (\ref{8e}), and (\ref{10e}),

\beq
\label{11e}
\phi '_0 \chi_0 - \phi_0 \chi ' _0 = \psi '_0 \chi_0 - \psi_0 \chi '_0 = 1 \ .
\eeq
\vskip 3 truemm

\noi {\bf Remark 3.} In general, $\phi_0 (0) \not= 0$, so that, the solution $\varphi_0 (r)$, with $\varphi_0 (0) = 0$, is a linear combination of the fundamental solutions $\phi_0$ and $\chi_0$~:

\beq
\label{12e}
\varphi_0 (r) = \alpha \phi_0 (r) + \beta_0 \chi_0 (r)\ .
\eeq

\noi Because of (\ref{7e}), $\varphi_0$ obviously satisfies (\ref{8e}). If we have $r^2V(r) \in L^1(1, \infty )$, then $\varphi_0 = \psi_0 + \alpha \chi_0$, and because of (\ref{9e}), $\varphi_0$ satisfies also (\ref{10e}). The solution $\psi_0$ being unique, the same is true for $\varphi_0$. Note that $\varphi_0 (r)$ is given by the solution of the Volterra integral equation (here we normalized it to $\varphi '_0 (0) = 1$~!)

\beq
\label{13e}
\varphi_0 (r) = r + \int_0^r (r-t) V(t) \varphi_0 (t) dt \ ,
\eeq

\noi whereas $\chi_0 (r)$ is the solution of

\beq
\label{14e}
\chi_0 (r) = 1 + \int_r^{\infty} (t-r) V(t) \chi_0 (t) dt \ .
\eeq
\vskip 5 truemm

In case $V(r)$ is positive, it is obvious on these equations, and on the basis of (\ref{6e}), that~:\\

a) $\varphi_0 (r)$ is an increasing convex function of $r$. Also $\varphi '_0 (r)$ is increasing, and therefore, $\varphi '_0 (r) \geq 1$ for all $r$. Assuming now also $r^2V(r) \in L^1 (1, \infty )$, one has

\beq
\label{15e}
\varphi ' _0 (\infty) = A < \infty \quad , \quad A = \int_0^{\infty} rV(r) \varphi_0 (r) dr\ ,
\eeq
\noi as seen on (\ref{13e}) ;\\

b) $\chi_0 (r)$ is a positive, convex, and decreasing function, with $\chi_0 (\infty ) = 1$. \\

\noi {\bf II - \underline{Proof of Theorem 1}} \\

The proof is based on the single formula for the phase-shift \cite{7r}

\beq
\label{16e}
\delta_0 (k) = - k \int_0^{\infty} {\varphi^2 (k, r) \over \varphi{'}^2 (k, r) + k^2 \varphi^2 (k, r)}\ V(r) dr \ ,
\eeq

\noi where one assumes $V(r) \geq 0$, $\varphi$ being the reduced radial wave function, solution of [1-4]

\beq
\label{17e}
\left \{ \begin{array}{l} \varphi '' (k, r) + k^2 \varphi (k, r) = V(r) \varphi (k, r)\ , \\ \\ r \in [0, \infty ) \ , \ \varphi (k, 0) = 0\ , \ \varphi ' (k, 0) = 1 \ .\end{array}\right .
\eeq
\vskip 3 truemm

\noi In some sense, (\ref{16e}) is an absolute definition of the phase-shift since, for potentials satisfying (\ref{2e}), one has, automatically, $\delta_0 (\infty ) = 0$. Note that, in (\ref{16e}), the normalization of $\varphi$ is irrelevant. However, we keep the normalization $\varphi ' (k, 0) = 1$ for convenience. The formula (\ref{16e}) is valid for all $k > 0$. One can prove, in fact, the following~: \\

\noi \underline{\bf Theorem 3 (\cite{7r}).} Under the conditions $V(r) \geq 0$ and $rV(r) \in L^1(0, \infty )$, the formula (\ref{16e}) is valid for all $k \geq 0$, $\delta_0 (k)$ is a continuous and bounded function, and $\delta_0( \infty ) = 0$. It is also differentiable for $k > 0$. In order to have also differentiability at $k = 0$, with a finite derivative $\delta ' (0)$, it is sufficient to have $r^2V(r) \in L^1(0, \infty )$.  In making $k \downarrow 0$ in (\ref{16e}), the integral diverges, according to Theorem~2 of Hille, if we only have $rV(r) \in L^1(0, \infty )$. However, there is the factor $k$ in front of it, and the net result is $\delta_0 (0) = 0$. \\

\noi {\bf Remark 4.} If we make $k = 0$ in (\ref{16e}), the denominator becomes $\varphi{'}^2(0, r)$. If there are bound states, $\varphi (0, r)$ has, according to the nodal theorem \cite{8r}, $n$ zeros in $(0, \infty )$. Between these zeros, it has maxima and minima, and so the integral is meaningless. In case of a resonance at $k=0$, one has $\varphi (0,  \infty )=$ constant, and $\varphi '(0,  \infty )=0$, and so, again, the breakdown of (\ref{16e}). If $V(r) \geq 0$, there are no bound states, and no resonance at $E = k^2 = 0$ \cite{1r,2r,3r,5r}.\\

\noi \underline{\bf Finiteness of ${\bf a}_{\bf 0}$}. Consider now 

\beq
\label{18e}
a_0 = \lim_{k\downarrow 0} {- \delta_0(k) \over k} \equiv - \lim_{k\downarrow 0}  {\delta_0(k) - \delta_0(0) \over k} = - \delta '(k=0)\  ,
\eeq

 \noi {\bf assuming V(r) $\geq$ 0}. Therefore, according to Theorem 3 above, $r^2V \in L(0 , \infty )$ is sufficient to secure that $\delta ' (0)$ is finite, that is, $a_0$ is finite. So, in essence, reference \cite{7r} contains already the proof of the old result that $r^2V \in L(0 , \infty ) \Rightarrow a_0$ finite. \par
 
 From (\ref{16e}), we also have 

\beq
\label{19e}
a_0 = \int_0^{\infty} V(r) {\varphi_0^2 (r) \over \varphi{'}_0^2 (r)}\ dr \ .
 \eeq
 
\noi  Again, on the basis of Theorem 2 of Hille, $r^2V \in L(0 , \infty )$ is also necessary to make $a_0$ finite. This completes the proof of the first part of Theorem 1. Finally, let us remark that, using $\varphi ''_0 = V \varphi_0$ in (\ref{19e}), and integrating by parts, we find $a_0 = \lim\limits_{R \to \infty} \left (R - {\varphi_0 (R) \over \varphi '_0 (R)}\right )$ also a known result.\\
 
  \noi \underline{\bf Finiteness of ${\bf r}_{\bf 0}$}. We must first compare (\ref{1e}) with (\ref{16e}). From the well-known expansion of cotg $x$~:
 
  \beq
\label{20e}
{\rm cotg}\ x = {1 \over x} - 2x \left ( {1 \over 6} + {x^2 \over 90} + {x^4 \over 945} + \cdots \right )\ ,
 \eeq
 
 \noi and using ($\delta_0$ is an odd function of $k$ [1-3])
 
  \beq
\label{21e}
\delta_0 (k) = n \pi - k a_0 + b k^3 + \cdots \ , \quad k \geq 0\ , 
 \eeq

\noi we find easily
 
  \beq
\label{22e}
r_0 = {2 \over 3} \ a_0 - {2b \over a_0^2}\ .
 \eeq
 
 \noi We assume now, of course, that $a_0$ is finite. It follows that the finiteness of $r_0$ and $b$ are completely equivalent. We can therefore concentrate ourselves on (\ref{16e}). Combining (\ref{16e}) and (\ref{19e}), we find 
 
  \bea
\label{23e}
&&b = \lim_{k \downarrow 0} {\delta_0 (k) + ka_0 \over k^3} = \lim_{k \downarrow 0}  {1 \over k^2} \nn \\
&&\int_0^{\infty} V(r) \ {\varphi^2(k,r) \ \varphi{'}_0^2(r) - \varphi_0^2(r) \ \varphi{'}^2(k,r) - k^2 \varphi_0^2 \varphi^2 \over \varphi{'}_0^2\left ( \varphi{'}^2 + k^2 \varphi^2\right )}\ dr\ .
 \eea
 
 \noi Writing now $\varphi^2 \varphi{'}^2_0 - \varphi_0^2\varphi{'}^2 = (\varphi \varphi{'}_0 - \varphi_0 \varphi{'}) (\varphi \varphi{'}_0 + \varphi_0 \varphi{'})$, we get, from (\ref{17e}) and $\varphi{''}_0 = V \varphi_0$, that $(\varphi \varphi{'}_0 - \varphi_0 \varphi{'})' = k^2  \varphi  \varphi_0$. Since, at $r= 0$, $ \varphi$ and $ \varphi_0$ vanish, and their derivatives are one (remember (\ref{2e}) !), we find 
 
  \beq
\label{24e}
\varphi(k,r)\  \varphi{'}_0(r) -  \varphi_0(r)\  \varphi{'}(k,r) = k^2 \int_0^r \varphi (k,t) \  \varphi_0(t) \ dt\ .
 \eeq
 
 \noi Therefore, taking the limit in (\ref{23e}), we find 
 
  \beq
\label{25e}
b = \int_0^{\infty} V(r) \left ( {2\varphi_0(r)\ \varphi{'}_0(r)  \int_0^r \varphi^2_0(t) \ dt - \varphi^4_0(r) \over \varphi{'}^4_0(r)}\right ) dr\ .
 \eeq
 
 \noi Now, from the asymptotic behaviour of $\varphi_0(r) \sim r$, and $\varphi{'}_(r)  \sim 1$, for $r \to \infty$, one sees immediately that the fraction in (\ref{25e}) behaves exactly as  $- {1 \over 3} r^4$. It follows that, for $V(r) \geq 0$, $b$ is finite if and only if $r^4 V(r) \in L^1(0, \infty )$. The same conclusions hold therefore for $r_0$. The net conclusion is~: \\
 
 \noi \underline{\bf Theorem 1$'$.} If $V(r) \geq 0$, $r^2V(r) \in L^1(0, \infty )$ is both necessary and sufficient for having $a_0$ finite. And for having also $r_0$ finite, it is both necessary and sufficient to have $r^4V(r) \in L^1(0, \infty )$. We must now include bound states.\\
 
 \noi \underline{\bf Bound states.} We shall choose, as is usually done, $\delta_0 (\infty ) = 0$. If there are bound states of energies $-\gamma_j^2$, $\gamma_j > 0$, $j = 1, \cdots n$, one has the Levinson Theorem (\ref{3e}) : $\delta (0) \equiv \delta (0) - \delta ( \infty ) = n \pi$. One can define then

 \beq
\label{26e}
\overline{\delta}_0 (k) = \delta_0 (k) - 2 \sum_{j=1}^n\  {\rm Arctg}\ {\gamma_j \over k}\ . 
 \eeq

\noi One has now, again, $\overline{\delta} ( \infty ) = 0, $ and $\overline{\delta}(0) = \delta (0) - n \pi = 0$. Using the inverse problem theory of Gel'fand and Levitan \cite{5r}, one can calculate the potential $\overline{V}(r)$ corresponding to $\overline{\delta} (k)$. If we write

 \beq
\label{27e}
V(r) = \overline{V}(r) + \Delta V(r) \ , 
\eeq

\noi it can be shown that the additional potential $\Delta V(r)$ has the following asymptotic behaviour \cite{5r}

 \beq
\label{28e}
\Delta V(r)  \ \mathrel{\mathop =_{r \to \infty}} \ - \sum C_j \ e^{-\gamma_j\cdot r}\ , 
\eeq

\noi where $C_j$ are positive constants. $\Delta V(r)$ keeps therefore a negative sign for large values of $r$, and is fast decreasing. We assume, of course, that $V(r)$ is not exponentially decreasing for, otherwise, there would be no problem, and $a_0$ and $r_0$ would be finite. So, for the purpose of Theorem 4, $V$ and $\overline{V}$ are equivalent. Since  $\overline{V}$ has no bound states, or a resonance at $E = 0$ because of $\overline{\delta}(0) = 0$, the theorem applies to $\overline{\delta} (k)$. Now, as it is easily seen by using the known expansion

 \beq
\label{29e}
{\rm Arctg}\ x \ \mathrel{\mathop =_{x \to \infty}} \ {\pi \over 2} - {1 \over x} + {1 \over 3x^3} + \cdots \ , 
\eeq

\noi the coefficients of the expansions of $\delta_0 (k)$ and $\overline{\delta}_0 (k)$ for small $k$, (\ref{21e}), are related to each other by (remember that, for $\overline{\delta}(k)$, $n=0$ in (\ref{21e}))

 \beq
\label{30e}
\left \{ \begin{array}{l} \overline{a}_0 = a_0 - 2 \displaystyle{\sum_j {1 \over \gamma_j}}\ , \\ \\  \overline{b} = b - 2 \displaystyle{\sum_j {1 \over \gamma_j^3}}\ . \end{array} \right . 
\eeq
 
\noi They differ by finite quantities, and this completes the proof of Theorem 1. \par

 If there is a resonance at $E = 0$, in the first term in the sum in (\ref{26e}) there is no factor 2, and one has $\delta (0) = {\pi \over 2} + (n-1)\pi$. In this case, as we said before (Remark~4), $\varphi_0 (\infty ) = C$, $\varphi '_0 (\infty ) = 0$, and so $a_0 = \infty$. \\
 
\noi {\bf III - \underline{Higher $\ell$}} \\ 
 
 The validity of the usual scattering theory leading to a continuous and bounded phase-shift $\delta_{\ell}(k)$ is, as we said in the introduction, secured always by $rV(r) \in L^1(0, \infty )$. Under this condition, the generalization of (\ref{16e}) for $\ell> 0$ is \cite{7r}~:
 
 \beq
\label{31e}
\delta_{\ell} (k) = - k \int_0^{\infty} {\varphi_{\ell}^2 (k, r) \over \left [ \left ( u'_{\ell} \ \varphi_{\ell} - u_{\ell} \ \varphi '_{\ell}\right )^2 + \left ( v'_{\ell} \varphi_{\ell} - v_{\ell} \varphi '_{\ell} \right )^2 \right ] }\ V(r)\ dr \ ,
 \eeq

\noi where $u_{\ell} (kr)$ and $v_{\ell}(kr)$ are appropriately normalized spherical Bessel and Neumann functions, and $\varphi_{\ell} (k, r)$ the solution of the reduced radial Schr\"odinger equation [1-4]

  \beq
\label{32e}
\left \{ \begin{array}{l} \varphi ''_{\ell} (k,r) + k^2 \varphi_{\ell} (k, r) = \left [ V(r) + \displaystyle{{\ell (\ell + 1) \over r^2}}\right ] \varphi_{\ell} (k,r) \ , \\ \\  \varphi_{\ell}(k, r)\  \displaystyle{ \mathrel{\mathop =_{r\to 0}} \ {r^{\ell + 1} \over (2 \ell + 1)!!}} + \cdots  \end{array} \right .
 \eeq
 
 \noi The Effective Range Formula becomes now [1-3]~:
  \beq
\label{33e}
k^{2\ell + 1} \ {\rm cotg}\ \delta_{\ell} (k) = - {1 \over a_{\ell}} + {1 \over 2}\ r_{\ell} k^2 + \cdots
 \eeq

In order to continue further, one needs now the equivalent of Theorem~2 in the presence of the centrifugal potential $\ell (\ell + 1)/r^2$. One can prove very easily, by mimiking the proofs ot Theorem~2 \cite{6r}, that one has~:\\

\noi \underline{\bf Theorem 2$'$.} Consider the equation
$$\phi ''_0(r) - V(r) \phi_0 (r) = {\ell (\ell + 1) \over r^2}\ \phi_0 (r)\ . \eqno(6')$$

\noi If $rV(r) \in L^1(0, \infty )$, this equation has a unique solution $\chi_0 (r)$ such that
$$\lim_{r \to \infty} r^{\ell} \chi_0 (r) = 1 \ , \quad \lim_{r \to \infty} \left ( r^{\ell} \chi_0 (r)\right ) ' = 0 \ . \eqno(7')$$

\noi If $V(r)$ is real, and ultimately keeps a constant sign, the condition on $V(r)$ is both necessary and sufficient for the existence of $\chi_0(r)$. There is also a solution $\phi_0(r)$, non-unique (modulo the addition of $\alpha \chi_0(r)$), such that  
$$\lim_{r \to \infty} r^{-\ell - 1} \phi_0 (r) = 1 \ , \quad \lim_{r \to \infty} \left ( r^{- \ell} \phi_0 (r)\right ) ' = 1 \ . \eqno(8')$$

\noi If $V(r)$ satisfies also the condition $r^2V(r) \in L^1 (0, \infty )$, then $r^{\ell} \chi_0 (r)$ satisfies
$$\lim_{r \to \infty} r\left [ r^{\ell} \chi_0 (r) - 1\right ] = 0 \ , \quad \lim_{r \to \infty} r^2\left ( r^{\ell} \chi_0 (r)\right ) ' = 0 \ , \eqno(9')$$

\noi and there exists a unique solution $\psi_0(r)$ such that
$$\lim_{r \to \infty} \left [ r^{-\ell} \psi_0 (r) - r\right ] = 0 \ , \quad \lim_{r \to \infty} r\left [ \left ( r^{-\ell} \psi_0 (r)\right ) ' -1 \right ] = 0 \ . \eqno(10')$$

\noi The solutions $\phi_0$ and $\chi_0$, or $\psi_0$ and $\chi_0$, are two independent solutions of ($6'$), and their wronskians are
$$\phi '_0 \chi_0 - \phi_0 \chi '_0 = \psi '_0 \chi_0 - \psi_0 \chi '_0 = (2 \ell + 1)\ . \eqno(11')$$

\noi Again, if $V(r)$ keeps a constant sign beyond some $R$, the condition $r^2V(r) \in L^1(1 , \infty )$ is both necessary and sufficient for the existence of $\psi_0$. In short, one gets Theorem $2'$ from Theorem 2 by replacing $\chi_0$ by $r^{\ell}\chi_0$, and $\phi_0$ and $\psi_0$ by $r^{-\ell}\phi_0$ and $r^{-\ell} \psi_0$, and changing the right-hand side of (\ref{11e}) to ($11'$). From these formulas ($7'$) to ($10'$), one can immediately obtain the asymptotic properties of $\chi '_0$, $\phi '_0$, and $\psi '_0$ themselves, to be used in (\ref{31e}).\par

The proofs are based on the Volterra integral equation for $\chi_0$ and $\phi_0$ \cite{1r,9r}~:

\beq
\label{34e}
\left \{ \begin{array}{l} \chi_0(r) = r^{-\ell} -  \displaystyle{\int_r^\infty {r^{\ell + 1} r{'}^{-\ell}- r^{-\ell} r{'}^{\ell + 1}\over (2 \ell + 1)}} V(r')  \chi_0 (r') dr' \ , \\ \\  \phi_0(r) = r^{\ell+1} -  \displaystyle{\int_r^\infty {r^{\ell + 1} r{'}^{-\ell}- r^{-\ell} r{'}^{\ell + 1}\over (2 \ell + 1)}} V(r')  \phi_0 (r') dr'  \ ,  \end{array} \right .
\eeq 

\noi by iterating them, starting from the zero order solutions $r^{-\ell}$ and $r^{\ell + 1}$, respectively, and mimiking exactly the proof of Theorem 2 of \cite{6r}. We leave the details to the reader. It is quite standard.\par

One can then continue the analysis, as was done for the $S$-wave, and one finds~: \\

\noi \underline{\bf Theorem 4.} In the absence of a bound state at $E=0$, if the potential keeps a constant sign for $r > R$, then~: a) for having a finite $a_{\ell}$, it is necessary and sufficient to have $r^{2\ell + 2}V(r) \in L^1(0 , \infty )$~; b) the effective range $r_{\ell}$ is finite if and only if $r^{2\ell + 4} V(r) \in L^1 (0 , \infty )$. In case there is a bound state at zero energy (for $\ell \geq 1$, it is a real bound state at $E=0$, with an $L^2$ wave function, instead of being a resonance. It contributes by $\pi$ to the Levinson Theorem), the scattering length $a_{\ell}$ is infinite [1-3], as in the case of a resonance when $\ell = 0$. This is obvious on (\ref{30e}), when making $\gamma_1 \downarrow 0$. \\

\noi {\bf Acknowledgement.} The author is grateful to Andr\'e Martin for having communicated to him his results prior to publication, and for useful correspondence. He wishes also to thank Professor Kenro Furutani, Reido and Takao Kobayashi, and the Science University of Tokyo (Noda) where this work began, for warm hospitality and financial support.

\end{document}